\documentclass[preprint,5p]{elsarticle}

\usepackage[utf8]{inputenc}
\usepackage{amsmath,graphicx}
\usepackage{amsfonts}
\usepackage{amssymb}
\journal{Chaos, Solitons and Fractals}
\begin{document}
\begin{frontmatter}

\title{Chaotic fluctuations in graphs with amplification}
\author[lepri1,mysecondaryaddress]{Stefano Lepri}
\address[lepri1]{Consiglio Nazionale delle Ricerche, Istituto dei Sistemi Complessi, Via Madonna del Piano 10 I-50019 Sesto Fiorentino, Italy} 
\address[mysecondaryaddress]{Istituto Nazionale di Fisica Nucleare, Sezione di Firenze, via G. Sansone 1 I-50019, Sesto Fiorentino, Italy}

\begin{abstract}
We consider a model for chaotic diffusion with amplification on graphs 
associated with
piecewise-linear maps of the interval. We investigate the possibility
of having power-law tails in the invariant measure by 
approximate solution of the Perron-Frobenius equation 
and discuss the connection with the generalized 
Lyapunov exponents $L(q)$. We then consider the case of 
open maps where trajectories escape and demonstrate that
stationary power-law distributions occur when $L(q)=r$,
with $r$ being the escape rate. 
The proposed system is a toy model for coupled active chaotic 
cavities or lasing networks and allows to elucidate 
in a simple mathematical framework the conditions for 
observing L\'evy statistical regimes and chaotic intermittency 
in such systems. 
\end{abstract}
\end{frontmatter}

\section{Introduction}

Dynamical systems defined on graphs are subject of 
current research, due to the many applications to model
complex interacting units with non uniform connectivity 
\citep{Porter}. Also, one of the features of complex systems is the 
possibility of display non-Gaussian fluctuations 
that make large rare events very relevant.
The distributions of the observables can have 
fat-tailed statistics leading to domination of a 
single event and lack of self-averaging of measurements.
In active systems where fluctuating amplification can occur, even rare 
trajectories can generate large-sized fluctuations. This is 
well-known for multiplicative stochastic processes \cite{Garcia-Ojalvo1999}
and chaotic dynamical systems
that display intermittency and multifractality 
\citep{crisanti2012products}. 

Countless example are present in the physical, biological 
and even social sciences. A case of experimental relevance is 
provided by optical media with diffusion and amplification of light, as 
it occurs in random lasers where 
heavy-tailed distributions of emission intensities, characterized 
by  L\'evy -stable statistics \cite{uchaikin1999} have been 
predicted \cite{Lepri2007} 
(see also \cite{Lepri2013,raposo2015analytical}) and confirmed 
in experiments \citep{Ignesti2013,Uppu2014,gomes2016observation}.

An experimental system that encompasses properties 
of a dynamical systems on graphs and non trivial statistics 
is the \textit{lasing network}, recently introduced
in \citep{lepri2017complex}. It consists of active and passive 
optical fibers, connected to form a graph structure. 
The connectivity induces a form of topological disorder and 
can be viewed as a discrete random laser, with a controllable 
complexity. The presence of the optical gain and disorder
induce wild emission fluctuations whose origin is not fully
understood  \citep{lepri2017complex}. 
So a related question is how such fluctuations relate 
to the network structure and connectivity.

The existence of fat tails is intimately related to the the possibility
for a spontaneous fluctuation to grow well beyond the average.
The indicators to quantify this are the finite-time and generalized 
Lyapunov exponents \cite{benzi1985characterisation}. 
For multiplicative noise they been shown  that they are useful tools to yield an intuitive derivation of the form of the probability distribution’s tail in the presence of additive noise \citep{Deutsch1993}.

Another possibility to have stationary fat-tailed 
statistics for multiplicative growing processes is to consider
resetting, namely a random process where the variable
is set to a given value with some given protocol \citep{manrubia1999stochastic}. The cases of stochastic
partial differential equations like 
the Kardar-Parisi-Zhang equation of fluctuating interfaces has been als considered \cite{gupta2014fluctuating}.

In the present paper we study a simple 
chaotic map that couples chaotic diffusion and random amplification.
Nonlinear maps are thoroughly investigated as mathematically
simple model to analyze 
the connection between macrolaws and microscopic chaos \cite{klages2007microscopic}. 
It can be regarded as a toy model for coupled active chaotic 
cavities or the lasing networks mentioned
above \citep{lepri2017complex}.  
The idea is that light rays can be treated as
particles undergoing chaotic diffusion and amplification.
Indeed, the classical dynamics of particles on graphs is a chaotic type of diffusive process \cite{barra2001classical}. Trajectories of 
a particle on a graph, undergoing scattering at its 
vertices, are in one-to-one correspondence  
with the ones of one-dimensional piecewise chaotic maps \cite{barra2001classical,pakonski2001classical}.

The model is simple enough to allow for a very detailed analysis,
demonstrating power-law distributions of the invariant measure.
It allows to elucidate 
in a simple mathematical framework the conditions for 
observing L\'evy statistical regimes and chaotic intermittency 
in such systems.
It also serves as an example to demonstrate the usefulness
of generalized Lyapunov exponents to assess the possibility
of power-law fluctuations. Moreover we extend the concepts 
to the case of open systems (like chaotic repellors), a case 
that, to our knowledge, has not been studied in this terms.

In Section \ref{sec:model} the map model is presented and
its relation with the physical systems is sketched. 
The stationary invariant measure is computed along with an effective
master equation. The connection between power-law and
generalized Lyapunov exponents is discussed in Section
\ref{sec:gle}. This relation is extended to the case of
open maps in Section \ref{sec:open}. 
The connection between the model and the calculation of 
the spectrum of the lasing network is given in the 
Appendix.

\section{Map model}
\label{sec:model}
We consider the following map
\begin{equation}
{\begin{cases} 
x_{n+1}=f(x_n) \\ E_{n+1}=g(x_n)E_n +s
\end{cases}}
\label{model}
\end{equation}
where $f$ is chaotic with a positive Lyapunov exponent $\lambda_1$.
For definiteness, let us consider $x$ to belong to the unit
interval and $E_n,g$  positive and $s\ge 0$ and small.

\begin{figure}[th] 
\includegraphics[width=0.45\textwidth]{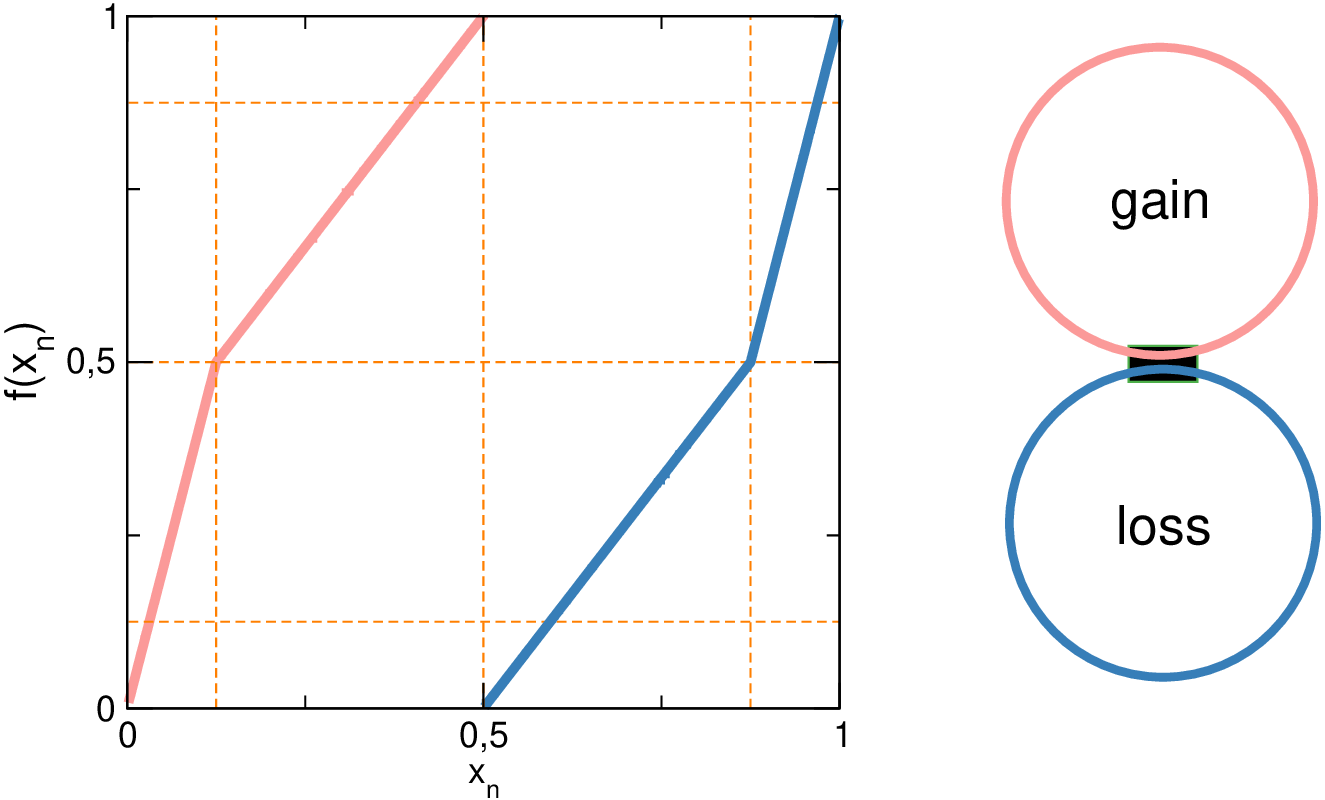} 
\caption{Left: the chaotic map (\ref{map}) for $p=0.25$.
Right: a sketch of two coupled chaotic cavities as in 
a double-ring lasing network
connected by a coupler that transmits with a given
probability (no reflections). One cavity contains an 
active amplifying medium, the other is dissipative. 
In the graph interpretation, the map $f$ can 
be derived exactly as a suitable Poincar\'e
section as done in \citep{barra2001classical}. 
}
\label{fig:map}
\end{figure}

As suggested by the notation, one can imagine $x_n$ to describe 
couples the position of a "ray" undergoing chaotic motion 
during which it acquires an "energy" that increases or 
decreases according to whether $g$ il larger or smaller than one.
Thus chaotic diffusion and amplification are coupled since the 
acquired energy depends on the trajectory. 
In Fig.\ref{fig:map} we sketch a physical reference system
inspired from the lasing network experimentally studied 
in \citep{lepri2017complex,giacomelli2019optical}.

The term $s$ represent some form of energy injection
and is needed to avoid that $E_n=0$ is not and "absorbing"
point.
The case $s=0$ leads to a non-stationary distribution
that for large times is log-normal. This is readily understood
as the variable $\log E_n =z_n$ performs a discrete-time
biased random walk \citep{fujisaka1986intermittency}.
The average velocity $\langle \log g(x_n)\rangle\equiv\lambda_2$
is the (second) Lyapunov exponent of (\ref{model}).   
The situation is drastically different in presence of a term $s> 0$,
that act as a source term. If $\lambda_2<0$ the variable
$z_n$ is attracted towards the source and this yield a 
stationary measure. In the stochastic case this 
mechanism of repulsion 
has been shown to generically yield power-law decaying
stationary distributions 
\citep{Sornette1997,nakao1998asymptotic, sornette1998multiplicative}.
Similar considerations apply for extended stochastic systems 
like the non-linear diffusion equation with multiplicative noise
\cite{munoz1998nonlinear}.
On the other hand, for $\lambda_2>0$ the variable grows 
and some form of saturating mechanism is needed to 
ensure unbounded motion (more on this below).

To keep the analysis as simple as possible we consider the case of the
piecewise-linear map
\begin{equation}
f(x)=
\begin{cases}
{1 \over p }x & 0\leq x\leq p/2\\
{1 \over 1-p}x+{1-2p \over 2(1-p)} & p/2<x\leq 1/2 \\ 
{1 \over 1-p}x-{1 \over 2(1-p)} & 1/2<x\leq 1-p/2 \\
{1 \over p}x +1 -1/p& 1-p/2<x\leq 1 
\end{cases}
\label{map}
\end{equation}
see Fig.\ref{fig:map}. If we consider the motion
of a particles on the graph
there drawn, $f$ can 
be derived exactly as a suitable Poincar\'e
section as described in \citep{barra2001classical}. 
\footnote{Thus the "physical" time $t_n$ corresponding to
the $n$th iteration of the map and depends
on the length on each bond . For instance, for
the graph in Fig.\ref{fig:map} with bond lengths
$L_1$ and $L_2$,  $t_{n+1}=t_n + T(x_n)$
 $T(x_n)=T_1$
or $T_2$ for $x_n<1/2$ and  $x_n>1/2$ respectively 
($v$ being the particle velocity, $T_{1,2}=L_{1,2}/v$).
This description can be generalized to arbitrary
graphs associated with Markov dynamics, see
\citep{barra2001classical}.}  
The map is everywhere expanding and is 
invariant for $x\to 1-x$, but it is straightforward
to generalize to an asymmetric case and/or more
complex graphs.

Since the invariant measure of the map 
is constant its the Lyapunov exponent 
$\lambda_1=-p\log p-(1-p)\log(1-p)$.
So $\lambda_1>0$ but it is vanishingly small for 
$p$ approaching 0 and 1 where the map has
weakly unstable orbits. 
Also, let us consider a piece-wise constant gain function
$g$
$$g(x_n)=\begin{cases} g & 0<x_n\leq\tfrac{1}{2}
\\ l & \tfrac{1}{2}<x_n<1
\end{cases}
$$ where $g\geq 1$ e $0<l\leq 1$. This is merely
a choice of simplicity and it entails that the 
sequence of multipliers $g(x_n)$ is in one-to-one
correspondence with symbolic dynamics of the map
$f$. Even in this simple example, for $p\neq 1/2$
the sequence of multipliers $g(x_n)$ is correlated
in time and the amplification fluctuations change
accordingly.

\begin{figure}[th]
\begin{center}
\includegraphics[width=0.4\textwidth]{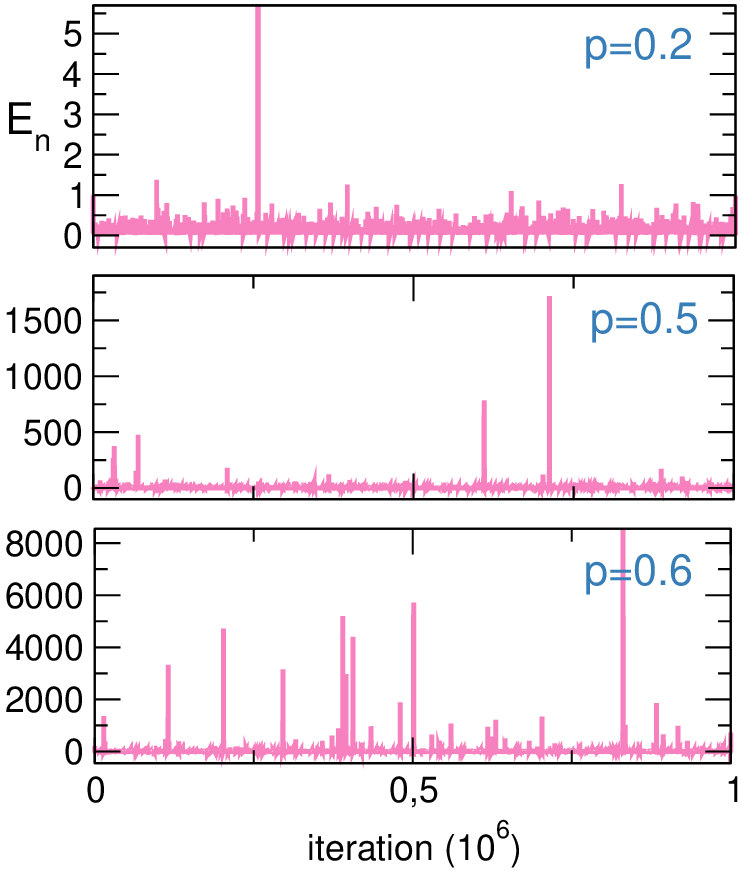} 
\end{center}
\caption{Time series of $E_n$, $g=1.2$ $l=0.8$ $s=10^{-3}$
for different values of $p$. The trajectory is highly intermittent
with large excursion of short duration.   
}
\label{fig:interm}
\end{figure}

Let us focus on the case where the solution does not diverge namely $\langle \log g(x)\rangle < 0$. Before entering the mathematical analysis, in 
Fig.\ref{fig:interm} we report some representative time-series
of the the variable $E_n$. The dynamics is highly intermittent
with large-amplitude spikes lasting tenths of iterates. 
A finite value of the source term insures that the variable 
does not vanish at long times, according to the mechanism mentioned 
above. 

We are interested in the statistics of the variable $E_n$.
The time evolution of the measure $P_n(x,E)$ can be
computed as solution of Perron-Frobenius operator 
\begin{align}
P_{n+1}(x,E)= \qquad \qquad  \qquad \qquad\qquad\\
=\begin{cases} {p \over
 g  }\;P_n(y_1,{E-s\over g  })+{(1-p) \over l}\;P_n(y_2,{E-s \over l }) & 0\leq x\leq {1 \over 2  } \\
 {(1-p) \over g  }\;P_n(y_3,{E-s\over g  })+{p \over
 l}\;P_n(y_4,{E-s\over l  }) & {1 \over 2  }< x\leq 1 \end{cases}
 \nonumber
\end{align}
where $y_1=px$, $y_2=(1-p)x+{1 \over 2  }$, $y_3=(1-p)x+{2p-1 \over 2}$
and $y_4=px+1-p$ are the preimages of $x$.

As a general approach, one may consider expanding $P_n$ on the 
basis of the eigen-functions of the Frobenius-Perron operator
of the map $f$. This would allow to describe the full evolution of 
the measure in time, including transients associated with 
possibly slow chaotic diffusion.
Since we are mostly interested in steady-state results, 
let us look for piecewise-constant in $x$ solutions of the 
form $P_{n}(x,E)=P_{1,n}(E)$ for $0<x<1/2$
and $P_{n}(x,E)=P_{2,n}(E)$ for $1/2\leq x<1$ respectively.
We obtain
\begin{align}
&P_{1,n+1}(E)=
\frac{p}{g} P_{1,n}\left(\frac{E-s}{g}\right)+
\frac{1-p}{l}{P_{2,n}\left(\frac{E-s}{l}\right) } \nonumber\\
&P_{2,n+1}(E)=
 {(1-p) \over g  }P_{1,n}\left({E-s\over g  }\right)
  +{p \over l}P_{2,n}\left({E-s\over l}\right)
  \label{master}
\end{align} 
We expect that this equation is valid when chaotic diffusion 
is sufficiently rapid to ensure homogeneization of the measure 
on the time scale faster than the typical growth.
It has a form of a master equation for probabilities 
of the variable $E$ on each side of the interval. 
Also, a standard Kramers-Moyal expansion may be used  to
show that it corresponds to a set of coupled Langevin
equations  with multiplicative noise 
for the energies on the two sides of the interval.
The connection between this equation and the one 
used in the calculation of the spectrum of the 
lasing network is given in the Appendix.

If we look for stationary solutions that decay 
as a power law for large $E\gg s$,
\begin{equation}
P_{1,2}(E) \propto A_{1,2} E^{-(1+\alpha)}
\label{power}
\end{equation}
where $A_{1,2}$ are constants, 
the dependence on $s$ can be neglected 
and we get the self-consistency condition
\begin{equation}
p(g^\alpha + l^\alpha)-(2p-1)g^\alpha l^\alpha =1
\label{alpha}
 \end{equation}
The latter, along with the the stability condition 
$\lambda_2<0$ i.e. $lg\le 1$ determines the region of 
validity of the power-law solution (\ref{alpha}). 
In particular, 
the more interesting case is when $\alpha<2$ yielding
distributions with diverging variance like 
in the well-know case of L\'evy stable distributions
\cite{uchaikin1999}.

Although the source term $s$ is essential to yield a 
stationary distribution, its value does enter in the 
exponent of asymptotic decay (\ref{power}).
We also checked numerically 
that basically the same statistics 
if found if $s$ is replaced by a random positive number
(additive noise).

\begin{figure}[th]
\begin{center}
\includegraphics[width=0.35\textwidth]{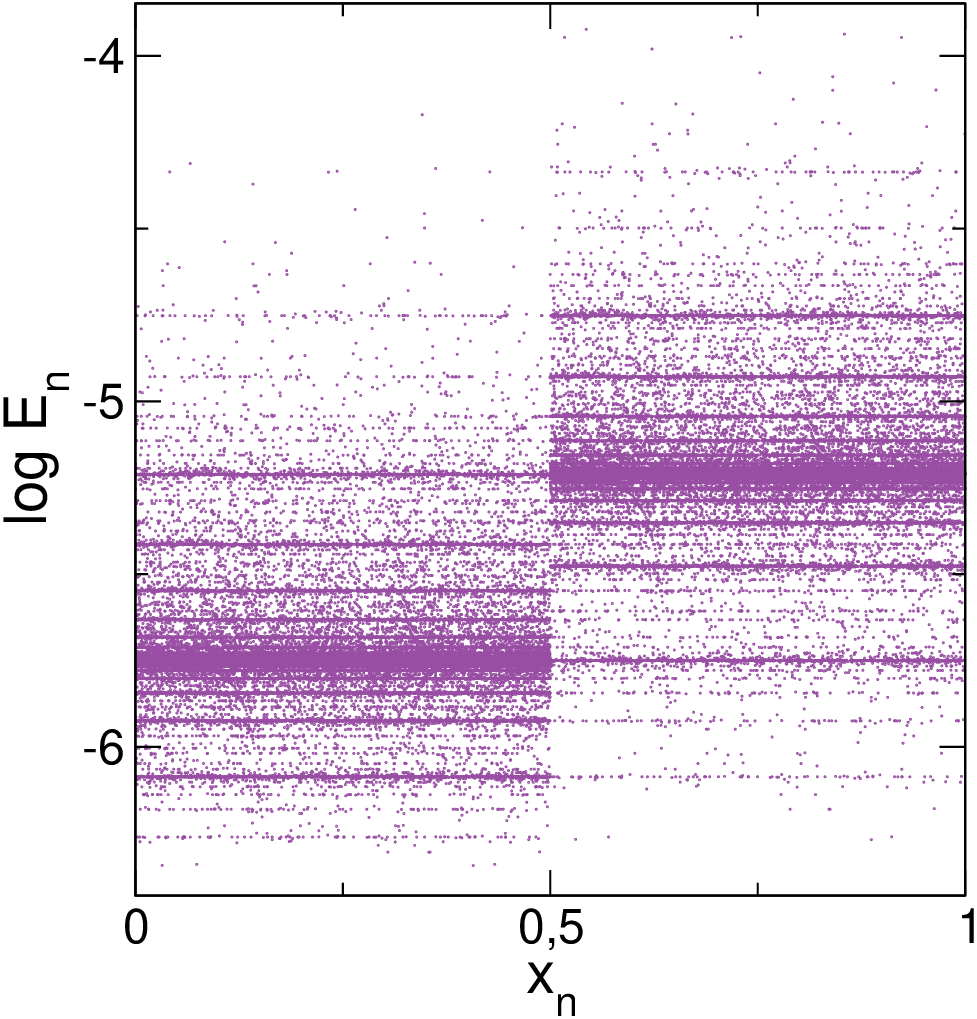} 
\end{center}
\caption{Iterates of the map  $x_n,\log E_n$, $g=1.4$ $l=0.4$ $s=10^{-3}$
for $p=0.05$, yielding $\lambda_1\approx 0.05$, $\lambda_2=-0.290$
giving a Lyapunov (Kaplan-Yorke) dimension $1.17$.   
}
\label{fig:fractal}
\end{figure}

To emphasize the importance of the separation 
of time scales leading to equations (\ref{master}) and thus
to power-laws let us compare with a situation when
chaotic diffusion is relatively slow. For instance, in
Fig.\ref{fig:fractal} we consider the 
case $p$ is small. 
In this limit, the map has a weakly unstable period two
orbit and $\lambda_1\approx p$.
Not surprisingly, the invariant measure is non uniform
and fractal in the direction of the variable $z$. 

\section{Generalized Lyapunov exponents}
\label{sec:gle}

The generalized Lyapunov exponents $L(q)$ define
the growth of the $q$th moment of the perturbation 
\citep{crisanti1988generalized,crisanti2012products,pikovsky2016lyapunov,vanneste2010estimating}. 
In general, for a perturbation $\delta u $ of a 
dynamical system, which evolves according to the linearized 
equation of motion, let $R(\tau) = \|\delta u(t+\tau)\|/\|\delta u(t)\|$
be the response function after a time $\tau$ to a disturbance at time $t$.
Then, for large times $\overline{R^q(\tau)}\sim\exp(L(q)\tau)$ 
where the overline denote a time average.
If $L(q) > 0$ for large enough $q$ then there is a finite
probability that a small perturbation grow very large.
Moreover, the deviation of $L(q)$ from a linear behavior
in $q$ signal an intermittent dynamics \cite{benzi1985characterisation}.

The condition for power-law stationary tails can be obtained
from generalized Lyapunov exponents \citep{Deutsch1993}.
In the present case we are interested in the generalized 
exponents associated with the $E_n$ variable 
when no source term is present and unbounded growth or 
decay at large times is possible. Actually, using 
equations (\ref{master}) with $s=0$ we can write the 
evolution map for the moments $\langle E_{1,2}^q\rangle$ 
and obtain $L$ as the logarithm of its largest eigenvalue,
\begin{equation}
L(q) = \log 
\left|\frac{p(g^q + l^q)+
\sqrt{p^2(g^q + l^q)^2-4(2p-1)g^q l^q} }
{2}\right|
\label{gle}
\end{equation}
In the simplest case $p=1/2$ multipliers are uncorrelated 
and one indeed gets the value 
$L(q) = \log \left(\frac{g^q + l^q}{2}\right)$ and
the moment multiplier is just the arithmetic average of $l^q$
and $g^q$. Note that, by construction, 
the standard Lyapunov exponent $\lambda_2=L'(q=0)=\log(gl)/2$
does not depend on $p$ while the $L(q)$ do. 

The stability condition implies that the Lyapunov exponent is 
negative. On the other hand, $L(q)\approx q \log g$ 
for $q$ large and positive. So a positive solution 
for $L(q_*)=0$ exist and coincides with  
the condition for a power-law decay given for 
to (\ref{alpha}) for $q_*=\alpha$. 
In the Gaussian approximation this is seen immediately
since in this case  
\[
L(q) \approx -|\lambda _2| q + \mu q^2
\]
where $\mu$ is the variance of $\lambda_2$. In this 
approximation $q_*\approx |\lambda _2|/\mu$ that makes transparent
the fact that fluctuations in the gain have to be of the same order
as $\lambda_2$ to observe large fluctuations.
This is in agreement
with the general scenario described in \citep{Deutsch1993}.

\begin{figure}[th]
\begin{center}
\includegraphics[width=0.4\textwidth]{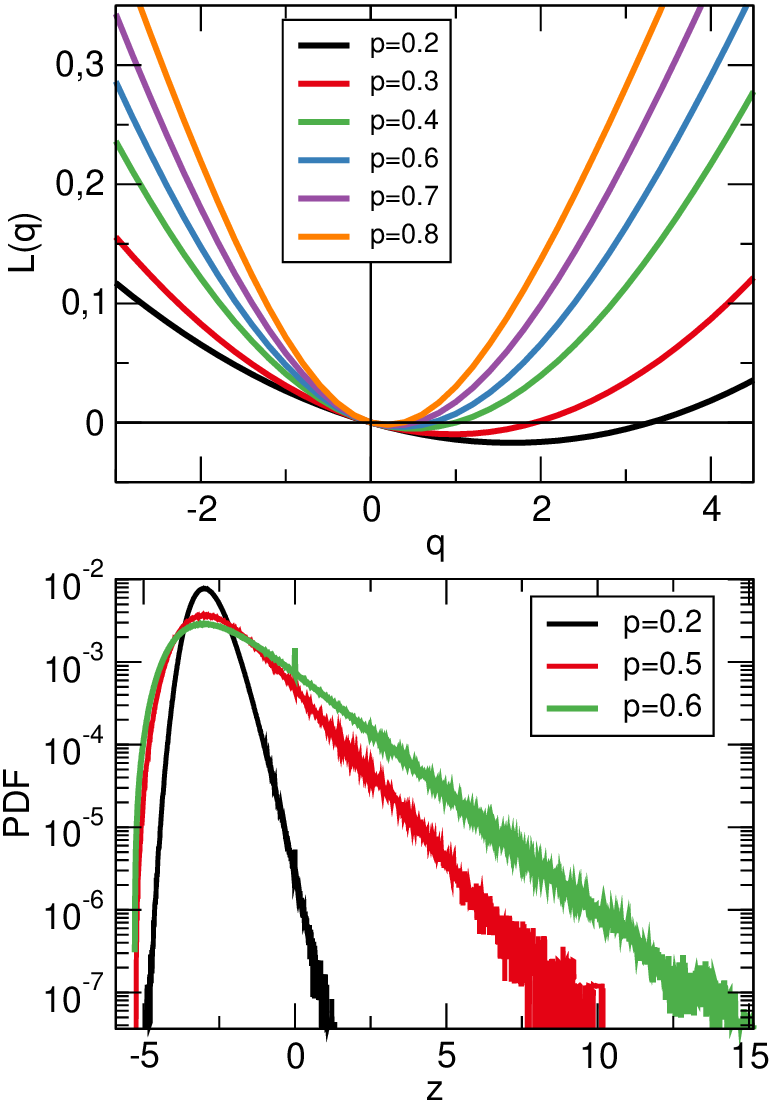} 
\end{center}
\caption{The generalized Lyapunov exponents $L(q)$
for different values of the map parameter $p$ and the 
distributions of $\log E$, $g=1.2$ $l=0.8$ $s=10^{-3}$.
An exponential tail of $P(z)\sim \exp(-\alpha z)$ correspond
to a power-law decay $E^{-1-\alpha}$.   
}
\label{fig:dmap}
\end{figure}

In Fig.\ref{fig:dmap} we report the generalized Lyapunov
exponents and the distributions of the $z_n$ variables.
The estimated exponents are in very good agreement with 
the simulations. For instance in the case $p=0.6$ the 
numerical histogram yield an exponent 0.66 to 
be compared with the value $q_*=0.71..$.

When the condition of stability is violated, some further mechanism 
of saturation is needed to have a steady distribution. In this 
case the term $s$ can be ignored. For instance one 
may consider a nonlinear term in the form of a 
chaotically-driven logistic map
\begin{equation}
{\begin{cases} 
x_{n+1}=f(x_n) \\ E_{n+1}=g(x_n)E_n - E_n^2 
\end{cases}}.
\label{modello}
\end{equation}
This a particular case of the systems thoroughly studied in 
\citep{fujisaka1986intermittency,yamada1986intermittency}
and will not thus be considered any further here.
In presence of the nonlinear term, $\lambda_2$ remains 
negative and the invariant measure remains broad but has 
an exponential cutoff at large $E_n$.
Detailed predictions on the nature of the chaotic intermittency close to the transition can be given \citep{fujisaka1986intermittency,yamada1986intermittency,fujisaka1987theory}, 
including universality of power spectra and Lyapunov exponents. 
In the next section, we consider an alternative possibility
for obtaining steady-state power laws even in the unstable case.

\section{Open maps}
\label{sec:open}

We now discuss another possibility to have steady fluctuations 
with power-law tails namely an open setup where the 
trajectory are allowed to escape (and be re-injected). The idea is that 
the distribution of the values of $E_n$ is in this case 
determined by the combined effect of the fluctuations 
of growth rates (as measured by finite-time Lyapunov 
exponents) and the statistical distribution of 
the escape events.

Let us consider the growth of a perturbation over a finite time $\tau$,
$E\propto \exp(\lambda(\tau)\tau)$ where $\lambda(\tau)$
is the finite-time Lyapunov exponent \citep{pikovsky2016lyapunov}.
Defining $z=\log E$ its distribution $\mathcal Q(z)$ is given by
\begin{equation}
\mathcal Q(z) = \int \int d\lambda d\tau \delta(z-\lambda(\tau)\tau) 
\mathcal P(\tau) \mathcal P(\lambda,\tau)
\label{qz}
\end{equation}
basically an average of the growth rates on the distribution 
of escape times $\mathcal P(\tau)$. As usual, for large $\tau$ we introduce the large-deviation function of the form
\begin{equation}
\mathcal P(\lambda,\tau) \sim \exp(-U(\lambda)\tau)
\end{equation}
In most cases, the distribution of escape times is Poissonian 
$\mathcal P(\tau) = r\exp(-r\tau)$ where $r$ is the escape rate.
Substituting this expression in equation (\ref{qz}), the 
resulting integral can be evaluated using the saddle-point 
approximation: if we denote 
by $\lambda_*$ the saddle point,  one obtains the 
condition $\lambda_* U'(\lambda_*)-U(\lambda_*)=r$. 
Then, recalling that
the generalized exponents are the Legendre transform 
of the large-deviation function $L(q)=q\lambda - U(\lambda)$
with $q=U'(\lambda)$ one obtains that the asymptotic
decay of the distribution
\begin{equation}
\mathcal Q(z) \sim \exp (-q_*z) ; \qquad L(q_*)=r
\label{qopen}
\end{equation}
Changing back to the original variable $E$ one obtains again
a power-law tail, $E^{-1-q_*}$ for large $E$. This 
last expression generalizes the one given above and 
confirms that also in the open setup
the generalized exponents can be used to estimate the 
power law decay. 

A consequence of the above is that, in the open case we can also 
consider the unstable case $\lambda_2 > 0$ and expect stationary
fat-tailed distributions. Indeed, the equation (\ref{qopen})
has a solution $q_*$ relatively close to zero. In the 
Gaussian approximation $L(q)=\lambda_2 q+ \mu q^2$ and for small $r$ 
one has $q_* \approx r/\lambda_2$. This nicely fits with the 
estimate given in \citep{Lepri2007} for observing 
L\'evy fluctuations in amplifying diffusive media with
absorbing boundaries, upon identifying $1/r$ with the
average residence time in the medium and $\lambda_2$
with the typical amplification time.

To verify the above argument we consider first the simpler
case of the map
(\ref{modello})  (with $s=0$) undergoing a stochastic resetting 
dynamics. With some preassigned small probability $r$ (which
represents the escape rate) 
the variable $E_n$ is reset to some arbitrary value (with no
modification on the $x_n$ dynamics). 

\begin{figure}[th] 
\includegraphics[width=0.45\textwidth]{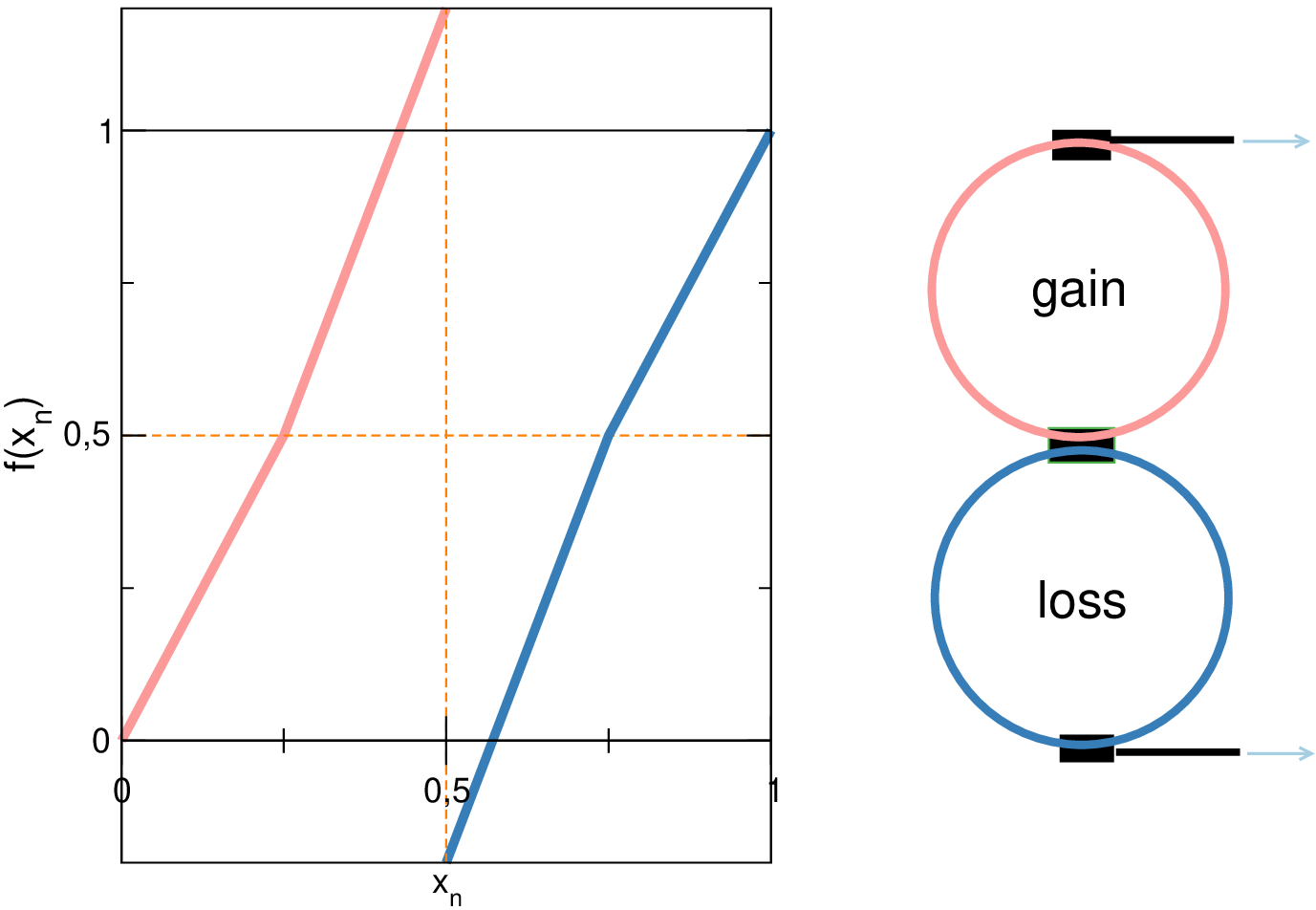} 
\caption{Left: the chaotic map (\ref{opmap}) for $a>2$.
Right: a sketch of the physical realization the 
open system in the case of two chaotic cavities 
connected by a coupler with leaks that allow escape of rays 
with some probability.
}
\label{fig:map2}
\end{figure}

The second case is a deterministic type of resetting, where
the $x_n$ dynamics is given by the map, see Fig.\ref{fig:map2}
\begin{equation}
f(x)=
\begin{cases}
2x & 0\leq x\leq 1/4\\
a(x-1/4)+{1 \over 2}& 1/4<x\leq 1/2 \\ 
a(x-3/4)+{1 \over 2} & 3/4<x\leq 1 \\
2x -1 & 3/4<x\leq 1 
\end{cases}
\label{opmap}
\end{equation}
The graph is in Fig.\ref{fig:map2} along with a sketch 
of the physical situation where particles are allowed 
to escape from the graph.
For $a>2$ there is escaping region in the 
interval $(1/4-1/2a,3/4-1/2a)$. Thus for $a\to 2^+$
the escape rate from the associated chaotic repellor is $r\approx (a-2)/4$.
Whenever the particle trajectory escapes from the 
unit interval, we reset $E_n$ to one and $x_n$ to a 
random uniformly-distributed value.

\begin{figure*}[th]
%\begin{center}
\centering
\includegraphics[width=0.85\textwidth]{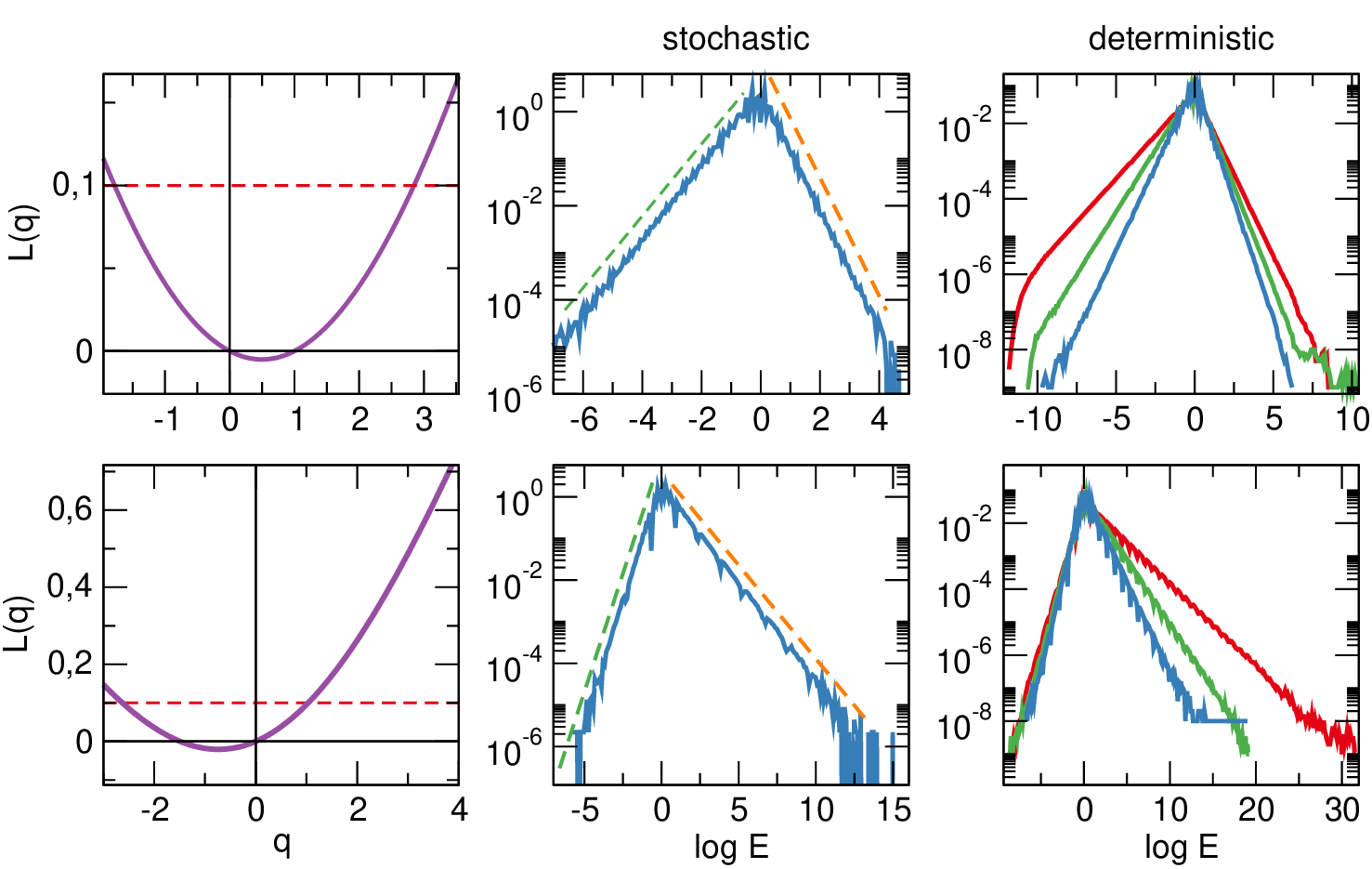}
%\end{center}
\caption{Leftmost panels: the generalized Lyapunov exponents
for $p=0.5$, $l=0.8$, the first row for the 
stable case  $g=1.2$, the second for the unstable one $g=1.4$.
Central panels: distributions of the variable $z=\log E$
for the map with stochastic resetting with probability $r=0.1$. Dashed
lines correspond to the exponential 
behaviors predicted by (\ref{qopen}).
Rightmost panels: distribution of the variable $z=\log E$
for the deterministic map (\ref{opmap}), 
$a=2.1, 2.2, 2.4$ (from top to bottom). The case
$a=2.4$ correspond to an escape rate    
$r\approx 0.083$.}
\label{fig:esc}
\end{figure*}

In Fig.\ref{fig:esc} we considered both examples in the 
unstable and unstable regimes. The distributions are 
clearly different from the one of the closed map given in
Fig.\ref{fig:dmap}. As expected, the deterministic and stochastic
case are similar.
The generalized Lyapunov exponents as given by formula 
(\ref{gle}) are also reported. In both cases, the distributions have 
double-exponential shape with rates in excellent agreement
with the one given by (\ref{qopen}), represented
graphically in the leftmost panels of Fig.\ref{fig:esc}. 
In the deterministic case with $a$ not too close from 2,
the escape rate has been estimated numerically.  

\begin{figure}[th]
%\begin{center}
\centering
\includegraphics[width=0.4\textwidth]{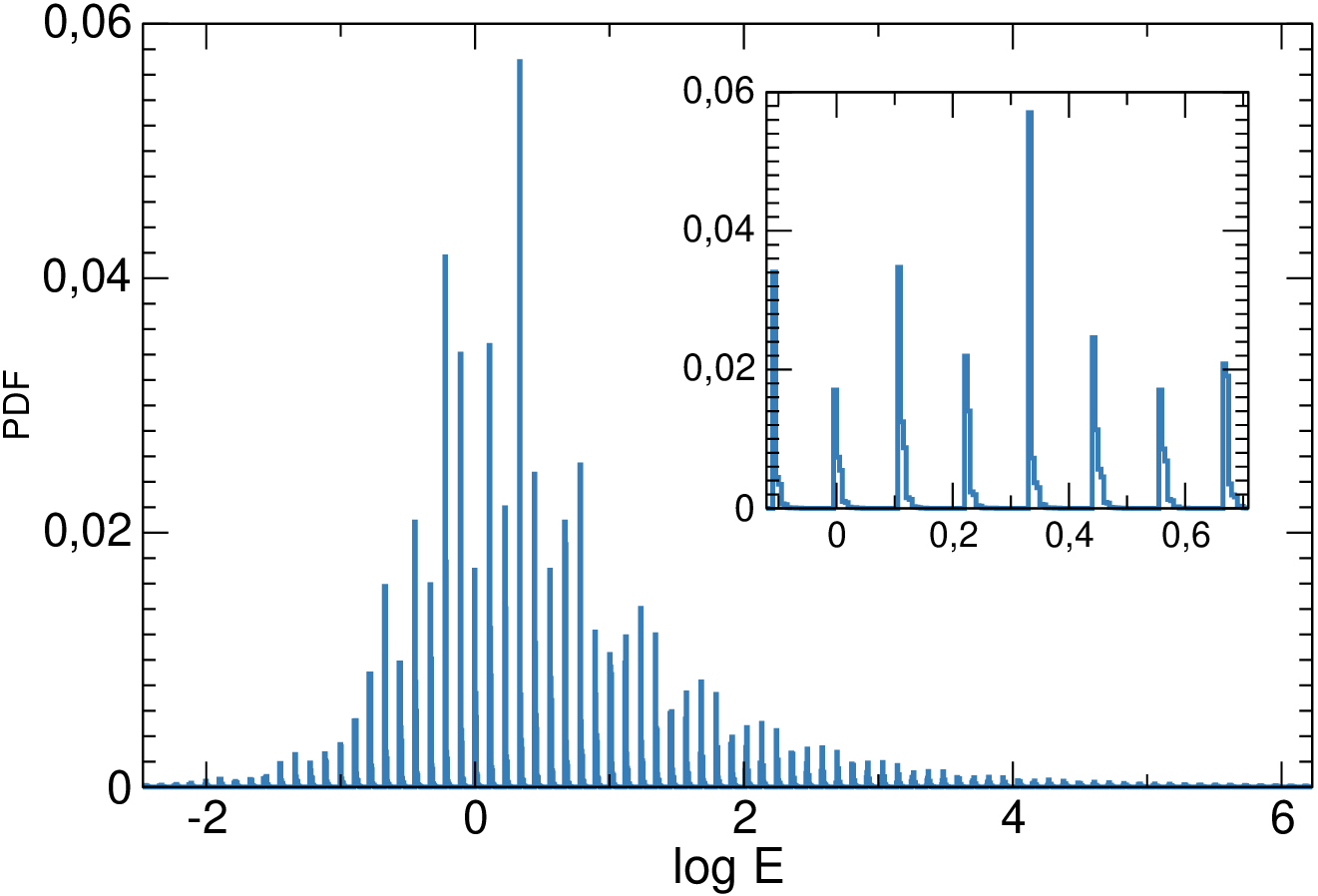}
%\end{center}
\caption{Finer structure of the distribution of $z=\log E$
for the open map (\ref{opmap}) with $a=2.4$ 
$l=0.8$, $g=1.4$ (unstable case). The inset shows an 
enlargement of the central part of the histogram.
The peaks are separated by a constant spacing approximatively
equal to $\log(gl)$.}
\label{fig:fhist}
\end{figure}

We conclude with a remark on the finer-scale structure
of the distribution.
A feature of the model is that the $z_n$ variable 
occurs almost in discrete values.
Fig. \ref{fig:fhist} shows that the distribution has a finer structure
with narrow peaks almost equally-spaced (see inset of 
Fig \ref{fig:fhist}).
That should be contrasted with the case of the closed map
where $z_n$ has continuous values and a smooth distribution.

\section{Conclusions}

Motivated by recent experiments of lasing networks \citep{lepri2017complex,giacomelli2019optical}, we have introduced
a toy map model describing the effect of chaotic diffusion and 
amplification on a graph structure. Since the motion 
is purely classical, it should apply when the wavelength 
is small with respect of the bond lengths. Evidence of large 
fluctuations and intermittency for the lasing
network has been indeed 
been provided both experimentally and by Monte-Carlo simulation
\citep{lepri2017complex}.  

Starting from a "Lagrangian" description in terms of chaotic 
trajectories we derived the corresponding "Eulerian" 
equations for the probabilities. We discussed the simplest
graph, but  the generalization to 
larger graphs is pretty straightforward especially at the 
level of the master equation (\ref{master}).
In this case the equation can be easily formulated in terms
of the transition matrix of the underlying diffusive 
process and the matrix for stochastic gain or loss
terms (see also the Appendix below).

Chaotic diffusion and amplification yield multiplicative
fluctuations and power-law steady distributions. In some 
regimes the variance can diverge leading to L\'evy-like 
statistics. We have confirmed that the Generalized Lyapunov exponents 
can give a precious hint on the statistics, both in the 
stable and unstable cases. We have extended this concept
to open systems through equation (\ref{qz}) that connects
the Generalized exponents with the escape rate. This result 
should apply under quit general conditions, 
as demonstrated by the case of random resetting
dynamics.

\section*{Acknowledgements}
I acknowledge Stefano Gelli for contributing to the initial stage 
of this work.

\section*{Appendix}

In \citep{lepri2017complex} the steady state modes of 
the lasing networks have been computed using an approach
extending the one used for quantum graphs \citep{kottos1999periodic}.
This is accomplished imposing that a suitable 
network matrix $N=SP$
has an eigenvalue equal to one.
Physically, $S$ is the scattering matrix of the optical couplers
(splitters) and $P$ is the so called propagation matrix
along the optical fibers and contains both the 
metric information on the bond length than the gain coefficients
\cite{giacomelli2019optical}.

To clarify the connection with 
the map model  studied here, let us first consider expressing equation (\ref{master}) 
in the "physical" time $t$
(neglecting the term $s$)
\begin{align*}
&P_{1}(E,t)=
\frac{p}{g} P_{1}\left(\frac{E}{g},t-T_1\right)+
\frac{1-p}{l}{P_{2}\left(\frac{E}{l},t-T_2\right) } \nonumber\\
&P_{2}(E,t)=
 {(1-p) \over g  }P_{1}\left({E\over g  },t-T_1\right)
  +{p \over l}P_{2}\left({E\over l},t-T_2\right)
  \label{master2}
\end{align*} 
where $T_{1,2}=L_{1,2}/v$ are the travel times (see the footnote in
the main text).
Taking the Laplace transform in $t$ and introducing the averages
\begin{equation*}
h(E,z) = \int_0^\infty h(E,t) e^{-zt}dt, \quad
I_{1,2} \equiv\int_0^\infty E P_{1,2}(E,z) dE
\end{equation*} 
one obtains the condition
\begin{equation*}
\begin{pmatrix}
 I_1 \\  I_2
\end{pmatrix} 
= W G\, \begin{pmatrix}
 I_1 \\  I_2
\end{pmatrix},
\end{equation*}
where
\begin{equation*}
W\equiv
\begin{pmatrix}
 p & 1-p \\  1-p & p
\end{pmatrix},\;
G\equiv 
\begin{pmatrix}
 g e^{zT_1} & 0 \\  0 & l e^{zT_2} 
\end{pmatrix} 
\end{equation*}
and $W$ is recognized to be the stochastic matrix for a random
walk on a graph with two states. To have non-trivial
solutions we impose $\det(WG-1)=0$ that determines 
all possible values of $z$.
This equation is obtained by the condition that $N$ has a 
eigenvalue one, by taking $|N|^2$, i.e.  the matrix
whose elements are the square moduli of it. 
The stochastic matrix of the graph is thus the square 
modulus on the scattering matrix $S$ of the coupler
$W=|S|^2$ while $G=|P|^2$.
Viewed in this way one can recognize the similarity 
with the "quantization" procedure outlined in 
\citep{tanner2000spectral,pakonski2001classical}
where the classical stochastic 
transition matrix is replaced by an unitary one
describing a quantum map.
The generalization of the above to arbitrary graphs
is straightforward.

\bibliography{mbiblio}

\end{document}